\documentclass[12pt,preprint]{aastex}

\def\kms  {km~s$^{-1}$}
\def\masy {mas~y$^{-1}$}

\def\etal {et al.~}
\def\eg   {e.g.~}
\def\ie   {i.e.~}

\def\Cyg  {Cygnus~X-1}
\def\Ro   {\ifmmode {R_0} \else {$R_0$} \fi}
\def\To   {\ifmmode {\Theta_0} \else {$\Theta_0$} \fi}

\def\mJypb{mJy~beam$^{-1}$}

\def\simless{\lower2pt\hbox{$\buildrel {\scriptstyle <}
   \over {\scriptstyle\sim}$}}
\def\pd  {\lower0pt\hbox{$\buildrel {^\circ} \over {.}$}}

\shorttitle{Trigonometric Parallax of \Cyg}
\shortauthors{Reid \etal}

\begin{document}

\title{The Trigonometric Parallax of \Cyg}

\author{Mark J. Reid\altaffilmark{1}, Jeffrey
        E. McClintock\altaffilmark{1}, Ramesh Narayan\altaffilmark{1},
        Lijun Gou\altaffilmark{1}, \\Ronald A. Remillard\altaffilmark{2},
        and Jerome A. Orosz\altaffilmark{3}}

\altaffiltext{1}{Harvard-Smithsonian Center for
   Astrophysics, 60 Garden Street, Cambridge, MA 02138, USA}
\altaffiltext{2}{Kavli Institute for Astrophysics and Space Research,
   Massachusetts Institute of Technology, 70 Vassar Street, Cambridge, MA 02139, USA}
\altaffiltext{3}{Department of Astronomy, San Diego State University,
   5500 Campanile Drive, San Diego, CA 92182, USA}

\begin{abstract}
We report a direct and accurate measurement of the distance
to the X-ray binary \Cyg, which contains the first black hole to be
discovered.  The distance of $1.86^{+0.12}_{-0.11}$~kpc was obtained from a 
trigonometric parallax measurement using the Very Long Baseline Array.  
The position measurements are also sensitive to the 5.6 d binary orbit and 
we determine the orbit to be clockwise on the sky.
We also measured the proper motion of \Cyg\ which, when coupled to
the distance and Doppler shift, gives the three-dimensional space motion of the system.
When corrected for differential Galactic rotation, the non-circular
(peculiar) motion of the binary is only about 21~\kms, indicating that
the binary did not experience a large ``kick'' at formation.
\end{abstract}

\keywords{astrometry --- black hole physics --- 
          stars: distances, individual (\objectname{\Cyg}) --- X-rays: binaries
         }

\section{Introduction}
In this paper (Paper I), we resolve the long-standing problem of the
distance to \Cyg\ by measuring its trigonometric parallax, ie,
triangulating using the Earth's orbit as one leg of the triangle and
measuring the change in its apparent position as the Earth orbits the
Sun.  Our results are based on observations of compact radio emission
with the National Radio Astronomy Observatory's\footnote{The National Radio 
Astronomy Observatory is a facility of 
the National Science Foundation operated under cooperative agreement
by Associated Universities, Inc.} 
Very Long Baseline Array (VLBA).  Previously, only one trigonometric parallax
has been published for a binary containing a black hole: V404~Cyg, an
X-ray binary with a low-mass ($M < 1 M_{\odot}$) companion
star~\citep{Miller-Jones:09}.  Our result is the first trigonometric parallax
for an X-ray binary with a massive companion ($M \gtrsim
10M_{\odot}$); an accurate distance is key to achieving secure and detailed 
models of the physical properties of the binary system.

\Cyg\ was the first black hole (BH) candidate to
be established via dynamical observations \citep{Webster:72,Bolton:72}. 
It is one of the most well-studied black hole systems with 
excellent optical and X-ray data.  However, modeling this system
is limited by its very uncertain distance.  Distance estimates span 
a very large range, with most  values between 1.8 and 2.4 kpc
\citep{Ziolkowski:05}.  The \Cyg\ distance estimate with the smallest 
{\it formal} uncertainty is $2.14\pm0.07$ kpc by \citet{Massey:95} 
and is based on photometry and spectroscopy of NGC 6871 (an open 
cluster in the Cyg OB3 complex).  While this distance estimate is very 
precise, its true accuracy (including systematic sources of error involving
extinction, metallicity corrections, and stellar modeling) is perhaps 
$\pm15$\%.  Also, it (and many other distance estimates) relies on the 
association of \Cyg\ with the Cyg OB3 complex.  While such an association is
supported by similar stellar (Hipparcos) and \Cyg\ (VLBI) proper motion 
measurements \citep{Lestrade:99,Mirabel:03}, it still must be considered tentative
as this line-of-sight through the Milky Way is very 
``crowded,'' passing nearly parallel to the Local spiral arm, 
which extends at least several kpc from the Sun, and then through the 
Perseus and Outer arms at greater distances \citep{Reid:09a}.

\section{Observations and Data Analysis}

Our observations were conducted under VLBA program BR141.  We observed 
\Cyg\ and two background sources over 10-hour tracks at five epochs:
2009 January 23, April 13, July 13, October 31 and 2010 January 25.
These dates well sample the peaks of the sinusoidal 
trigonometric parallax signature in both Right Ascension and Declination.
This sampling provides near maximum sensitivity for parallax detection
and ensures that we can separate the secular proper motion 
(caused by projections of Galactic rotation, as well as any peculiar 
motion of \Cyg\ and the Sun) from the sinusoidal parallax effect.
Table~\ref{table:positions} lists the positions of \Cyg\ and the 
background continuum sources, as well as their angular separations 
from \Cyg.

\begin{deluxetable}{lllrr}
\tablecolumns{5} \tablewidth{0pc} \tablecaption{Source Positions}
\tablehead {
  \colhead{Source} & \colhead{R.A. (J2000)} &  \colhead{Dec. (J2000)} &
  \colhead{$\theta_{\rm sep}$} & \colhead{P.A.} 
\\
  \colhead{}       & \colhead{(h~~m~~s)} &  \colhead{(d~~'~~'')} &
  \colhead{(deg)} & \colhead{(deg)}
            }
\startdata
 \Cyg ............  &19 58 21.672762 &35 12 05.72511 & ... & ...   \\
 J1953+3537 ...     &19 53 30.875712 &35 37 59.35927 & 1.1 & $-66$ \\
 J1957+3338 ...     &19 57 40.549923 &33 38 27.94339 & 1.6 & $-175$\\
\enddata
\tablecomments {  The position for \Cyg\ at 8.45 GHz comes from the parallax fitting
  at epoch 2009.570.  This position is relative to the unweighted
  average of the two background 
  sources, which gave coordinate differences of about $\pm0.25$ mas,
  consistent with their ICRF catalog uncertainties of $0.5$ mas \citep{Fey:04}.
  Angular offsets ($\theta_{\rm sep}$) and position angles (P.A.) east of north 
  relative to \Cyg\ are indicated in columns 4 and 5. 
  }
\label{table:positions} 
\end{deluxetable}

Generally, data calibration followed similar procedures as for 
parallax observations of continuum sources in the Orion nebular cluster 
at 8.4 GHz \citep{Menten:07}.  
We placed observations of well-known strong sources 
near the beginning, middle and end of the observations in order to 
monitor delay and electronic phase differences among the intermediate
frequency bands.
In practice, however, we found minimal drifts and used only a single
scan of J2005+7752 for this calibration.  

In order to calibrate atmospheric propagation delays, we scheduled
``geodetic blocks'' consisting of typically a dozen ICRF sources \citep{Fey:04}
observed over a wide range of source azimuths and elevations 
as describe in \citet{Reid:09b}.  We observed in left circular polarization
and spread eight 16-MHz bands so as to span $\approx500$ MHz of bandwidth.
We corrected the interferometer data for ionospheric delays using
global models of the total electron content obtained from global positioning
system observations \citep{Walker:00}.   Residual multi-band delays and fringe rates were 
measured and used as data to solve for clock offsets and drift rates and 
unmodeled zenith atmospheric propagation path delays.  Later these were
removed from the rapid-switching (parallax) data.

The rapid-switching observations employed four adjacent frequency bands 
of 16-MHz bandwidth and recorded both right and left circularly polarized
signals.  The four (dual-polarized) bands were centered at frequencies
of 8.425, 8.441, 8.457 and 8.473 GHz.
We alternated between two $\approx11$~min blocks,
each consisting of observations of \Cyg\ and one of the background
sources.  Within a block, we switched sources every 40~s, 
typically achieving 30~s of on-source data.  We used the background
sources J1953+3537 and J1957+3338 for phase-referencing.

We calibrated the correlated data using the NRAO Astronomical Image 
Processing System (AIPS).  The calibration sequence included four steps. 
The first step involved correction of interferometer delays and phases for
the effects of diurnal feed rotation (parallactic angle),
for errors in the values of the Earth's orientation parameters
used at the time of correlation, and for any small position shifts 
in the {\it a priori} coordinates of either the maser or background sources.
Since the VLBA correlator model includes no ionospheric delays, we used 
global total electron content models to remove ionospheric effects.
At this point, we also corrected the data for residual zenith atmospheric 
delays and clock drifts (determined from the geodetic block data).  

In the second step, we adjusted interferometer visibility amplitudes for 
the small (few \%) effects of biases in the threshold levels of the data
samplers at each station.  We also entered system temperature and
antenna gain curve information into calibration tables, which allows
conversion of correlation coefficients to flux density units.
In the third step, we performed a ``manual phase-calibration'' to remove
delay and phase differences among all eight 16-MHz bands.
This was done by selecting one scan on a strong calibrator, fitting
fringe patterns to the data for each frequency band, and then shifting 
delays and phases to remove offsets.  
The fourth calibration step involved determining antenna-based
complex gains from each of the two background sources and
interpolating and removing those from the \Cyg\ observations.  
For most antennas at most times the phases were easily ``connected.''
However,  when interferometer phase differences between adjacent 
background source scans exceeded $60^\circ$,  
the data between those times were discarded.  

\begin{figure}
\epsscale{1.0} 
\plotone{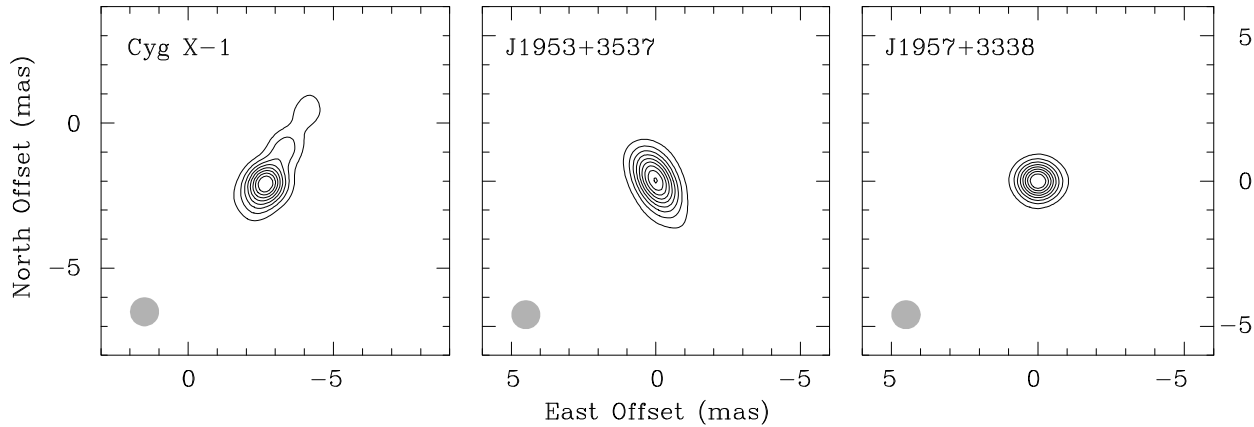} 
\caption{
  Images of \Cyg\ and the two background continuum sources from
  the middle epoch (2009 July 13). Source names are in the upper left
  corner and restoring beams are in the lower left corner
  of each panel.  All contour levels are integer multiples of
  1 \mJypb\ for \Cyg\ and 15 \mJypb\ for the background sources. 
  Zero contours are suppressed for clarity; negative brightness levels
  were below the first contour level for all images.
  \label{images}
        }
\end{figure}

We then imaged the calibrated data with the AIPS task IMAGR.
The synthesized ``dirty'' beam typically had a full-width at 
half-maximum of $1.55\times0.86$ mas 
elongated approximately in the north-south direction.  We adopted a
round convolving-beam of 1.0 mas to construct CLEAN images.  
Example images from the middle epoch are shown in Fig.~\ref{images}.
Positions were determined by fitting elliptical Gaussian brightness 
distributions to the images.  For \Cyg, which shows a core-jet 
structure, we restricted the fitted pixels to those close to the
bright core, typically within 1~mas of the peak emission.
The peak brightness of \Cyg\ ranged between 4 and 9 mJy beam$^{-1}$
among our observations.

\section {Parallax and Proper Motion Fitting}

\begin{figure}
\epsscale{1.0}
\plotone{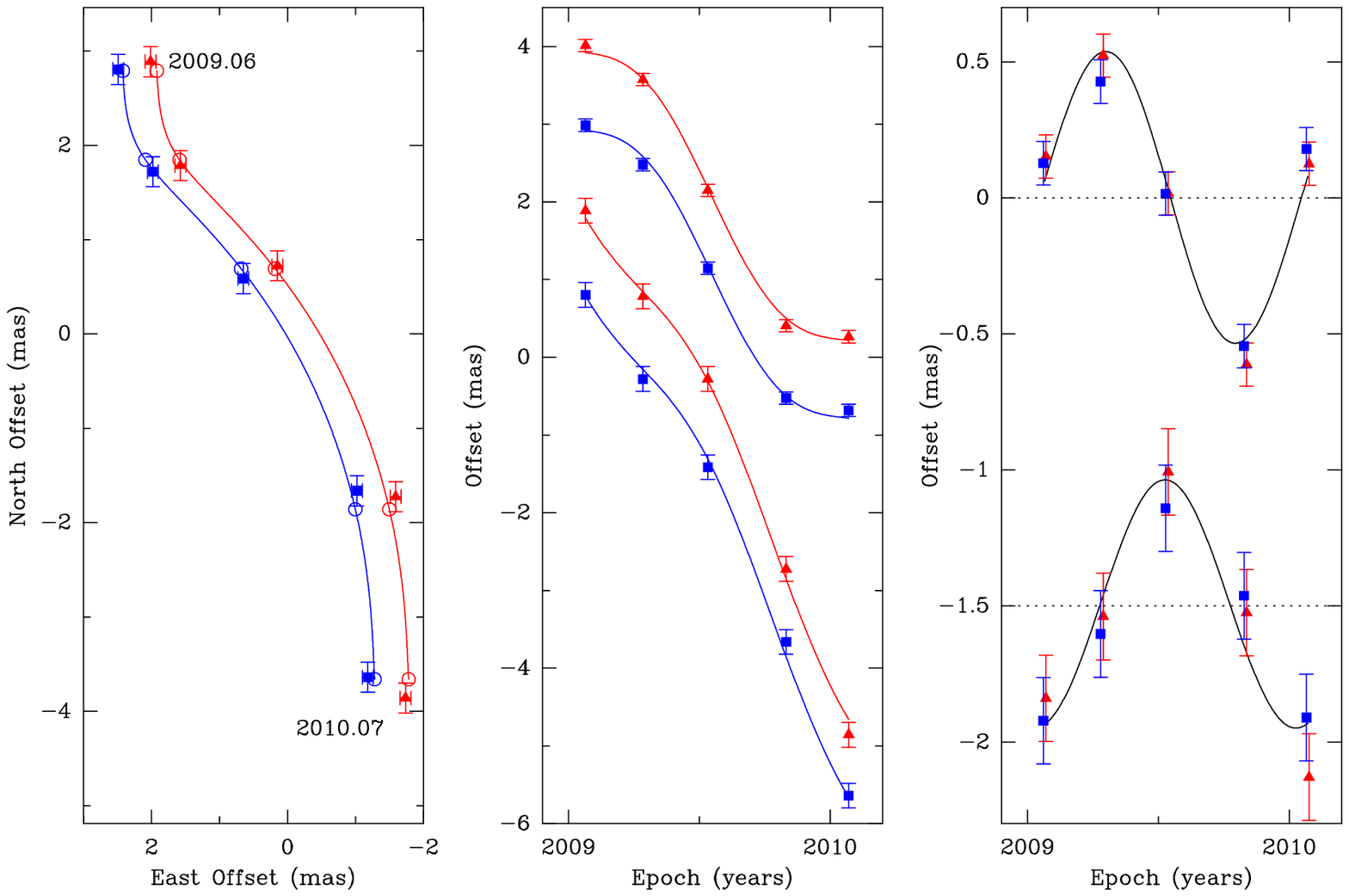} 
\caption{
  Parallax and proper motion data and fits.
  Plotted are position offsets of \Cyg\ relative 
  to the two background sources: J1953+3537 {\it (red triangles)} and
  J1957+3338 {\it (blue squares)}.  The degree-scale separations
  of the absolute positions given in Table~\ref{table:positions} have been removed.
  The yearly sinusoidal position variations (the parallax 
  signature) caused apparent position shifts of \Cyg\ as the Earth 
  orbits the Sun.
  {\it Left Panel:} Positions on the sky with first and last epochs labeled.
  Data for the two background source are offset horizontally for clarity.
  The expected positions from the parallax and proper motion fit
  are indicated {\it (circles)}.
  {\it Middle Panel:} east {\it (top lines)} and North {\it (bottom lines)} 
  position offsets and best fit parallax and proper motions fits versus time.  
  Data for the two background sources and two coordinates are offset vertically
  for clarity.
  {\it Right Panel:} Same as the {\it middle panel}, except the
  with the best fit proper motions have been removed, allowing
  all data to be overlaid and the effects of only the parallax seen.
  The north offset data have been offset vertically (below) the
  east offset data for clarity.
  \label{parallax_fits}
        }
\end{figure}

The change in position of \Cyg\ relative to a background 
continuum source was then modeled by the parallax
sinusoid in both coordinates, completely determined by one parameter
(the parallax), and a secular proper motion in each coordinate
(see Fig.~\ref{parallax_fits}).
The model included the effects of the ellipticity of the Earth's orbit.
The weighting of the data in the parallax and proper motion 
fitting is complicated because the formal position uncertainties 
are often unrealistically small, since {\it a priori} unknown sources 
of systematic error often dominate over random noise.
The north-south components of relative positions often
have greater uncertainty than the east-west components, because
the interferometer beams are generally larger in the north-south direction and 
systematic errors from unmodeled atmospheric delays usually
are more strongly correlated with north-south positions \citep{Honma:08}.  

In order to allow for, and estimate the magnitude of, systematic errors, 
we assigned independent ``error floors'' to the east-west and north-south position 
data and added these floors in quadrature with the formal position-fitting 
uncertainties.  Trial parallax and proper motion fits were conducted and 
a reduced $\chi^2_\nu$ (per degree of freedom) statistic was 
calculated separately for the east-west and north-south residuals.  
The error floors were then adjusted iteratively so as to make 
$\chi^2_\nu\approx1.0$ for each coordinate.  This procedure resulted in error floors of
0.08 and 0.16~mas for the eastward and northward position measurements, 
respectively.  The magnitudes of these error floors are reasonably
consistent with those obtained for other parallax targets observed
at 8.4 GHz, \eg \citet{Menten:07}, and are probably dominated by
unmodeled ionospheric delays.  Any component of variability in the
centroid of the core-jet position of \Cyg\  caused by changing
jet opacity must be less than $\approx0.1$~mas.

\begin{deluxetable}{lllllrr}
\tablecolumns{7} \tablewidth{0pc} 
\tablecaption{\Cyg\ Parallax \& Proper Motion Fits}
\tablehead {
  \colhead{\Cyg\ Orbit} & \colhead{Background} &  
  \colhead{Parallax} & \colhead{$\mu_x$} &
  \colhead{$\mu_y$} &\colhead{$\chi^2$} & \colhead{N}
\\
  \colhead{}      & \colhead{Source} & 
  \colhead{(mas)} & \colhead{(\masy)} &
  \colhead{(\masy)} &    & \colhead{}
            }
\startdata
 None~~~ ........& J1953+3537    &$0.582\pm0.055$ &$-3.74\pm0.10$ &$-6.56\pm0.19$ & 4.57 & 5 \\
 ~~~~~~~~~~ ........& J1957+3338 &$0.511\pm0.061$ &$-3.66\pm0.11$ &$-6.28\pm0.21$ & 5.54 & 5 \\
 ~~~~~~~~~~~........& Combined   &$0.547\pm0.041$ &$-3.70\pm0.08$ &$-6.42\pm0.14$ &12.82 &13 \\
\\
 CCW ~~~........ & Combined      &$0.528\pm0.047$ &$-3.68\pm0.09$ &$-6.36\pm0.17$ &16.81 &12 \\
 CW ~~~~~........& Combined      &$0.539\pm0.033$ &$-3.78\pm0.06$ &$-6.40\pm0.12$ & 8.49 &12 \\
\enddata
\tablecomments {Combined fits used a common parallax and proper motion
parameters for both background sources.  The proper motion components are
defined as $\mu_x = \mu_{\alpha}\cos{\delta}$ and $\mu_y = \mu_\delta$, 
where $\alpha$ is Right Ascension and $\delta$ is Declination. The binary orbital model assumed
an orbital period of 5.599829~d, superior conjunction at JD 2441874.71~d, 
a radius of the radio source about the center of mass of 0.071~mas, 
an inclination of $36^\circ$, and spin-axis of $-26^\circ$ east of north.
We tried two orbital direction: clockwise (CW) and counter-clockwise (CCW) on the sky.
Columns 6 and 7 list the post-fit $\chi^2$ values and the number of
degrees of freedom ($N$).  Parameter uncertainties have been re-scaled to those 
appropriate for a $\chi^2_\nu$ (per degree of freedom) of unity.
               }
\label{table:fits}
\end{deluxetable}

The fitting results are presented in Table~\ref{table:fits}.  
First, we attempted separate parallax fits for each background source 
measurement.  The parallaxes for \Cyg\ for the two background sources
differ by only 0.071~mas and are consistent with their individual formal uncertainties 
of about 0.06~mas.  This agreement suggests that neither atmospheric 
mis-modeling nor potential unresolved structural variations in the background 
sources contribute position errors greater than the error floors. 

Next, because of the agreement of the \Cyg\ parallaxes and proper motions 
for the two background sources, we did a combined fit with common parallax
and proper motion parameters for both sets of data.  (In the combined fit,
we did allow for different position offsets for \Cyg\ with respect to each
background source, since the {\it a priori} positions for these sources were not
known to better than $\approx0.5$~mas.)  

Finally, we investigated the effect of the \Cyg\ binary orbit on the 
parallax fits.  All orbital parameters are known with reasonable accuracy 
and were considered as independent variables \citep{Brocksopp:99,Gies:08,Orosz:11}, 
with the exception of the direction of the orbital rotation on the sky. 
The binary orbital model we adopted had
an orbital period of 5.599829~d, superior conjunction at JD 2441874.707~d,  
a radius of the radio source about the center of mass of 0.128~AU, 
an inclination of $36^\circ$ (ie, the angle between the orbit angular 
momentum vector and our line-of-sight).  (Using the revised value of $i=27^\circ$, 
determined in Paper II, results in no significant difference in parallax and 
proper motion fits.) 
We assumed an orbital angular momentum vector parallel to the radio jet in \Cyg, 
observed in our images (see Fig.~\ref{images}) and previous studies 
\citep{Stirling:01} to be about $-26^\circ$ east of north.  This leaves two 
possible orbital directions, clockwise (CW) and counterclockwise (CCW) on the sky, 
to be determined by the data.

\begin{figure}
\epsscale{1.0}
\plotone{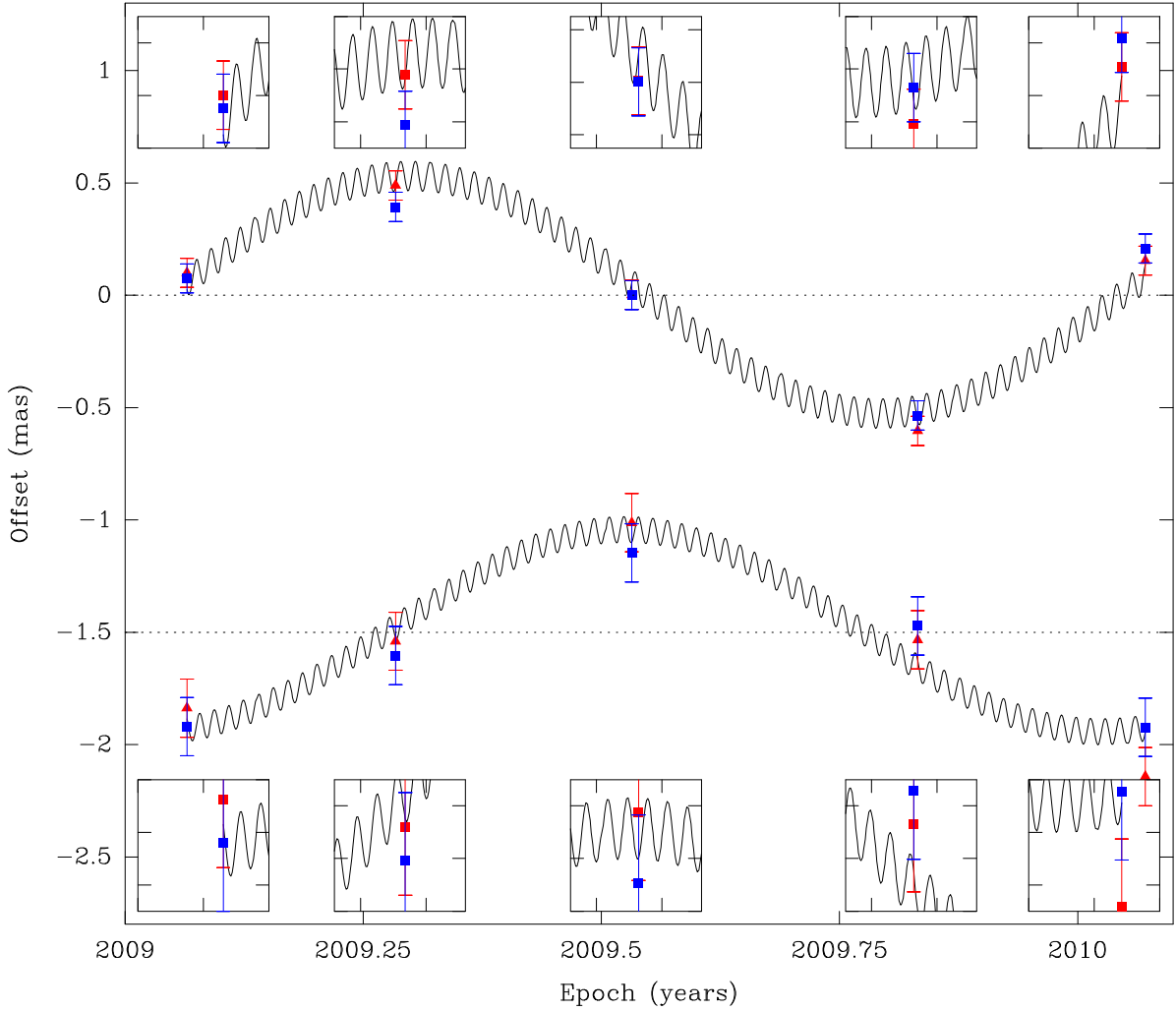} 
\caption{
  Parallax fit including the effects of a clock-wise (on the sky) binary 
  orbit for the radio source relative to the center of mass of \Cyg.  
  East {\it (top)} and north {\it (bottom)} position offsets 
  and best fitting model {\it (lines)} versus time.  
  Plotted are the positions of \Cyg\ relative to the two background sources: 
  J1953+3537 {\it (red triangles)} and J1957+3338 {\it (blue squares)}.
  Clearly seen is the annual parallax sinusoidal signature (with the
  proper motion removed) and the ``high frequency'' 
  oscillations caused by the 5.6 day orbit of the radio source (the black hole) 
  about center of mass of the binary system.  Blow-ups of the eastward 
  {\it (top small boxes)} and northward {\it (bottom small boxes)} offsets
  each cover 0.1 year in time and 0.25~mas in offset.
  \label{fit_with_orbit}
        }
\end{figure}

Parallax and proper motion fits for both orbital directions are listed in 
Table~\ref{table:fits}.  The $\chi^2$ value for the CW orbit was 8.49 
(for 13 degrees of freedom), which was significantly lower than the value of 16.81 
for the CCW orbit.  The agreement of the data with the CW orbit model
can be seen in the position data in Fig.~\ref{fit_with_orbit}, especially in the 
more accurate eastward offset data.  Thus, our astrometric observations 
appear to be sensitive to the position changes caused by orbit of the radio source 
(presumably the black hole) about the center of mass of the binary system, and we 
have directly determined the direction of the orbit to be clock-wise on the sky.

Adopting the CW orbit, we re-fitted the data allowing the orbital radius 
to vary and found a radius of $0.18\pm0.09$~AU.  Thus, the current VLBA
data independently gives a $2\sigma$ estimate of the radius of the compact
radio source about the center of mass of the binary.   With more observations 
in the future, we could refine the values of some of the \Cyg\ orbital 
parameters.   

We adopt a \Cyg\ parallax of $0.539 \pm 0.033$~mas from the fit with 
the lowest $\chi^2_\nu$ value.  That fit combined the data from both
background sources and used the orbital model with clockwise 
projected motion on the sky.  
This parallax for \Cyg\ corresponds to a distance of $1.86^{+0.12}_{-0.11}$~kpc.  
At this distance, the proper motion of \Cyg\ corresponds to 
$-33$~\kms\ and $-56$~\kms\ eastward and northward, respectively.  
Completing the full space velocity, the average heliocentric radial
velocity of \Cyg\ is about $-5$~\kms\ \citep{Gies:08}.

\section{Discussion}

Since we have measured the position, parallax, proper motion, and others
have measured the line-of-sight velocity, we have full three-dimensional
position and velocity information for \Cyg.  We can convert this velocity 
vector from the equatorial heliocentric reference frame in which they are 
measured into a Galactic reference frame.  A convenient Galactic frame is one
rotating with a circular velocity $\Theta(R)$ at the position of the source: 
ie, a ``local standard of rest'' at the location of \Cyg.  This transformation 
is described in detail in \citet{Reid:09a}.

Conversion to a rotating Galactic reference frame depends somewhat upon the
solar motion and Galactic parameters.   We adopt the distance of the Sun 
from the Galactic center of $\Ro= 8.25$ kpc as a rounded average of the
values quoted by \citet{Reid:93} for a variety of ``classical'' techniques
and by \citet{Ghez:08} and \citet{Gillessen:09} from the orbits of stars
about the supermassive black hole at the Galactic center.  The rotation speed 
of the local standard of rest is $\To \approx 240$ \kms, which is based on 
this distance and the proper motion of Sgr~A* (the super-massive black hole 
at the Galactic center), which directly measures 
$(\To+V_\odot)/\Ro=30.2$~\kms~kpc$^{-1}$ \citep{Reid:04}. 
Using the newly proposed Solar motion of $V_\odot \approx 11$~\kms\  
by \citet{Schonrich:10}, and assuming a locally flat rotation curve, 
\ie\ $\Theta(R) = \To$,  we find a ``peculiar motion'' (relative to circular 
motion) for \Cyg\ of $17\pm1$~\kms\ toward the Galactic center, $-11\pm3$~\kms\ 
in the direction of Galactic rotation, and $6\pm1$~\kms\ toward the North 
Galactic Pole.  
Thus, the total magnitude of the peculiar motion of \Cyg\ is only 21~\kms, 
indicating that the binary system did not experience a large ``kick'' when 
the black hole formed.  This result supports the conjecture that the black hole 
may have formed without a supernova explosion~\citep{Katz:75,Mirabel:03}.

Recently, \citet{Xiang:11} modeled the scattering by interstellar
dust grains of x-rays from \Cyg.   They considered 18 models describing the
characteristics of dust and selected six models with iron column densities
consistent with a value of $1.62\times10^{17}$~cm$^{-2}$ (Lee et al., in preparation).  
Modeling the x-ray light-curves, three of the six yielded 
distances between 3.15 and 3.59 kpc, and these models were discarded as being 
incompatible with our trigonometric parallax.  The remaining three models yielded 
distances between 1.72 and 1.90 kpc, compatible with our parallax.  
Thus, having an accurate trigonometic parallax can be important 
for the study of x-ray scattering from time varying sources.

The black hole in \Cyg\ is fed gas via the stellar wind emanating
from its O-type companion, which underfills its Roche equipotential
lobe.  The degree to which the star underfills its Roche lobe must be
tightly constrained in order to obtain a useful dynamical model of the
system, and accurate knowledge of the source distance is necessary to
accomplish this~\citep{Orosz:07}.  Using the accurate distance, $D$,
reported here, we have constructed such a model and obtained precise
values for the mass, $M$, of the black hole, the orbital inclination
angle, $i$, and other system parameters, as reported in Paper
II~\citep{Orosz:11}.  With the required, accurate values of $M$, $i$ and
$D$ in hand, we proceed in Paper III~\citep{Gou:11} to model the X-ray
continuum spectrum and determine the spin of the black hole.  BH spin can 
be determined because the measured X-ray spectrum from the accretion disk 
is dominated by radiation coming from very near the BH.
The temperature and luminosity of the X-ray emission
increase greatly with increasing BH spin, as the inner radius
of the accretion disk, $R_{\rm in}$, shrinks toward the BH event horizon.  
$R_{\rm in}$ can be determined by fitting the X-ray continuum 
emission from the BH's accretion disk to a fully relativistic model 
if the black hole mass, distance and accretion disk inclination angle
are known \citep{Li:05}.  Because
an astronomical black hole (which is believed to have negligible net
electric charge) is completely described by specifying its mass and
spin, our measurements of these quantities provide a {\it complete}
description of the black hole in \Cyg.

\vskip 1.0truecm\noindent
Facilities: VLBA

\end{document}